\documentclass[aps,prb,twocolumn,showpacs,superscriptaddress, longbibliography, floatfix]{revtex4-2}

\usepackage{xcolor, graphicx}
\usepackage{amssymb,amsmath, amsfonts, bm}
\usepackage{newtxtext,newtxmath}
\usepackage{float}
\usepackage{dcolumn}% Align table columns on decimal point
\usepackage[none]{hyphenat} % avoid the hyphen "-" that splits a word at the end of a line.
\usepackage{tabularx}
\usepackage{multirow} % 用于合并行
\usepackage{booktabs}
\usepackage{braket}
\renewcommand{\arraystretch}{1.5}

\usepackage{multirow}
\usepackage{hyperref}
\hypersetup{
    colorlinks=true,
    linkcolor=blue,    
    urlcolor=blue,
    citecolor=blue,
    }
    
\usepackage{multibib}
\newcites{SI}{Reference}
\usepackage{easyReview}

\begin{document}
\title{Resilient cluster Mott states in layered Nb$_3$Cl$_8$ against pressure-induced symmetry breaking}

\author{Hongbin Qu}
\affiliation{School of Physical Science and Technology, ShanghaiTech University, Shanghai 201210, China}

\author{Xiaoqun Wang}
\author{Hai-Qing Lin}
\affiliation{School of Physics, Zhejiang University, Hangzhou 310027, Zhejiang, China}
\author{Gang Li}
\email{ligang@shanghaitech.edu.cn}
\affiliation{School of Physical Science and Technology, ShanghaiTech University, Shanghai 201210, China}
\affiliation{\mbox{ShanghaiTech Laboratory for Topological Physics, ShanghaiTech University, Shanghai 201210, China}}
\affiliation{State Key Laboratory of Quantum Functional Materials, ShanghaiTech University, Shanghai 201210, China}

\date{\today}

\begin{abstract}

In this work, by combining density-functional theory (DFT) with dynamical mean-field calculations (DMFT), we compare the crystal and electronic structures of the prototype cluster Mott insulator Nb$_{3}$Cl$_8$ at ambient and high-pressure. 
We explain the finite but significantly reduced charge gap experimentally observed at $P=9.7$ GPa.
We reveal a local symmetry breaking of the Nb$_{3}$ trimer under pressure, reducing its symmetry from $C_{3v}$ to $C_{s}$. 
This leads to a strong bandwidth enhancement and a lift of band degeneracy.  
Crucially, despite the significant change of band details, the cluster Mott insulating state is robust against local symmetry breaking. 
We show that the experimental observed gap under pressure is still a cluster Mott gap and its reduced value stems from both increased bandwidth and reduced Coulomb interactions under pressure. 
Our study provides the first systematic theoretical elucidation of how pressure-induced symmetry breaking dictates the cluster Mott state, establishing a robust foundation for understanding the intricate relationship between symmetry, local/non-local correlations, and emergent quantum states in correlated cluster compounds.

\end{abstract}

\maketitle

\section{Introduction}

Mott insulator is one of the hallmark of condensed matter physics which differs it from other disciplines of physics by the emergent electron localization from many-body interactions~\cite{hubbard_electron_1963, mott_metal-insulator_1968, doi:10.1126/science.177.4047.393, dagotto_complexity_2005}. 
The Mott gap triggered by local Coulomb interactions freezes the mobility of electrons yielding an insulating ground state for bands with commensurate electron occupancy~\cite{mott_metal-insulator_1968, imada_metal-insulator_1998}. 
In Mott insulators, the competition of Coulomb energy ($U$) and bandwidth ($W$) drives the metal-insulator transition (MIT) that governs the ground state of many transition metal compounds, such as NiO, MnO, V$_2$O$_3$, $\kappa$-(ET)$_{2}$Cu$_{2}$(CN)$_{3}$ET, etc~\cite{2003Sci_302_89L, 1969PhRvL.23.1384M, 1970PhRvB...2.3734M, PhysRevB.7.1920, 1980PhRvB..22.2626K,doi:10.1126/science.1088386, PhysRevB.44.1530, PhysRevLett.53.2339, PhysRevB.44.6090, 2003PhRvL..91j7001S, 2011ARCMP...2..167K, 2017RvMP...89b5003Z}. 
In conventional Mott insulators, the electrons are localized at atomic site such that electron hopping is strictly forbidden. Thus, there is no long-range transport contributing to conductance. 
However, the suppression of long-range conductance does not require electrons to be strictly localized. 
They can still freely move in a short-range scale but having no long-range transportation.  
Such states are called cluster Mott insulators. 
Electrons are delocalized inside a small cluster but transportation between clusters are significantly suppressed. 
Typical cluster Mott insulators include Mo$_3$O$_8$ family~\cite{cotton_metal_1964, mccarroll_structural_1977, cotton_preparations_1991, sheckelton_possible_2012, flint_emergent_2013, mourigal_molecular_2014, chen_cluster_2016, chen_emergent_2018}, two-dimensional transition metal dichalcogenides~\cite{TaSe2_cluster, 2010Natur.464..199B, TaS2-nat.m, TaSe2-nat.p,tas2-Law2017,tas2-He2018,tas2-Shi2021,tas2-Shen2022,tas2-Ruan2021,tas2-Tian2024,tas2-Wang2020,tas2-Manas-Valero2021,tas2-Klanjsek2017}, and $AM_4X_8$ ($A$=Ga, Ge, Al; $M$=V, Ti, MO, Nb, Ta; $X$=S, Se, Te)~\cite{GaVSe_2000, PhysRevLett.93.126403, https://doi.org/10.1002/pssb.202100160, PhysRevB.99.100401, Lee_2019, PhysRevB.102.081105, PhysRevResearch.2.022017, Petersen_2023}.
Some of them are also conjectured as good candidates of quantum spin liquids (QSL)~\cite{law_1t-tas2_2017,zhang_enhanced_2024}. 

Recently, van der Waals layered material family Nb$_3$X$_8$ (X = Cl, Br, I) has been proposed as a new promising material platform for studying the cluster Mott insulating state~\cite{khvorykh_niobium_1995,kennedy_chemical_1991, kennedy_experimental_1996,sheckelton_rearrangement_2017, haraguchi_magneticnonmagnetic_2017, miller_solid_1995, pasco_tunable_2019, nanolett_feng_2022, hu_correlated_2023, grytsiuk_nb3cl8_2024, gao_discovery_2023,  aretz_strong_2025, liu_direct_2025}. 
Their simple structure and high quality of single crystal make this family of materials ideal for understanding the difference of cluster Mott insulators to the conventional ones.  
The unique in-plane breathing mode makes the atomic wave functions of three Nb atoms highly entangled forming a molecular state. 
Meanwhile, the specific valence configuration of Nb leaves one unpaired electron in this molecular state yielding a half-filled molecular orbital, which turns into a Mott insulator under electron Coulomb interactions. 
The entangled nature of this molecular state significantly enhances non-local Coulomb interactions among the atomic states, which can be as strong as the local ones. 
The emergence of strong non-local Coulomb interactions is the characteristic of cluster Mott insulators. 

Compared to the much discussed Mott insulators~\cite{RevModPhys.78.17, https://doi.org/10.1002/qute.202200186, 23-20231508, Mott_spinrtronics}, there is less known about the electrical and structural response of the cluster Mott states. 
Recently, high pressure experiments  have been conducted~\cite{jiang_pressure-driven_2022,shan_pressure-induced_2023,hong_structural_2025}, providing valuable information on structure engineering of Nb$_{3}$Cl$_{8}$.  
A continuous evolution from insulating to metallic states was observed~\cite{shan_pressure-induced_2023,hong_structural_2025}. 
Unfortunately, due to the lack of detailed structure information in the high-pressure metallic states~\cite{shan_pressure-induced_2023,hong_structural_2025}, it is not clear whether the metallization of Nb$_{3}$Cl$_{8}$ is induced by structural transition or not. 
As it is widely known, even without structural transition, pressure alone can trigger insulator-metal transition by modifying the bond length, crystal-field splitting, etc. 
Furthermore, by shrinking crystal volume with pressure, the kinetic energy is enhanced and, effectively, the Coulomb interaction between electrons is reduced. 
The interaction-driven Mott insulating state is then less favored at high pressure, resulting in the insulator to metal transition. 

In this work, based on the experimentally available structure data at $P=9.7$ GPa~\cite{jiang_pressure-driven_2022}, we theoretically compare the structural and electronic evolution of Nb$_3$Cl$_8$ at low-pressure (LP) and high-pressure (HP). 
We reveal that a local symmetry breaking induced by external pressure is in charge of the electronic structure change and the size of charge gap, excellently explaining the experimental observations~\cite{jiang_pressure-driven_2022}. 
We find that, while the local symmetry breaking enhances the bandwidth and reduces the effective Coulomb interaction strength, the cluster Mott insulating state survives at HP. 
Our study provides the first systematic theoretical elucidation of the electronic structure evolution in cluster Mott insulators under pressure-induced symmetry breaking. 
These findings establish a robust theoretical foundation for understanding the intricate relationships between symmetry, local and non-local electronic correlations, and emergent quantum states in transition metal cluster compounds.

\section{results}

\begin{figure}
\includegraphics[width=1.0\linewidth]{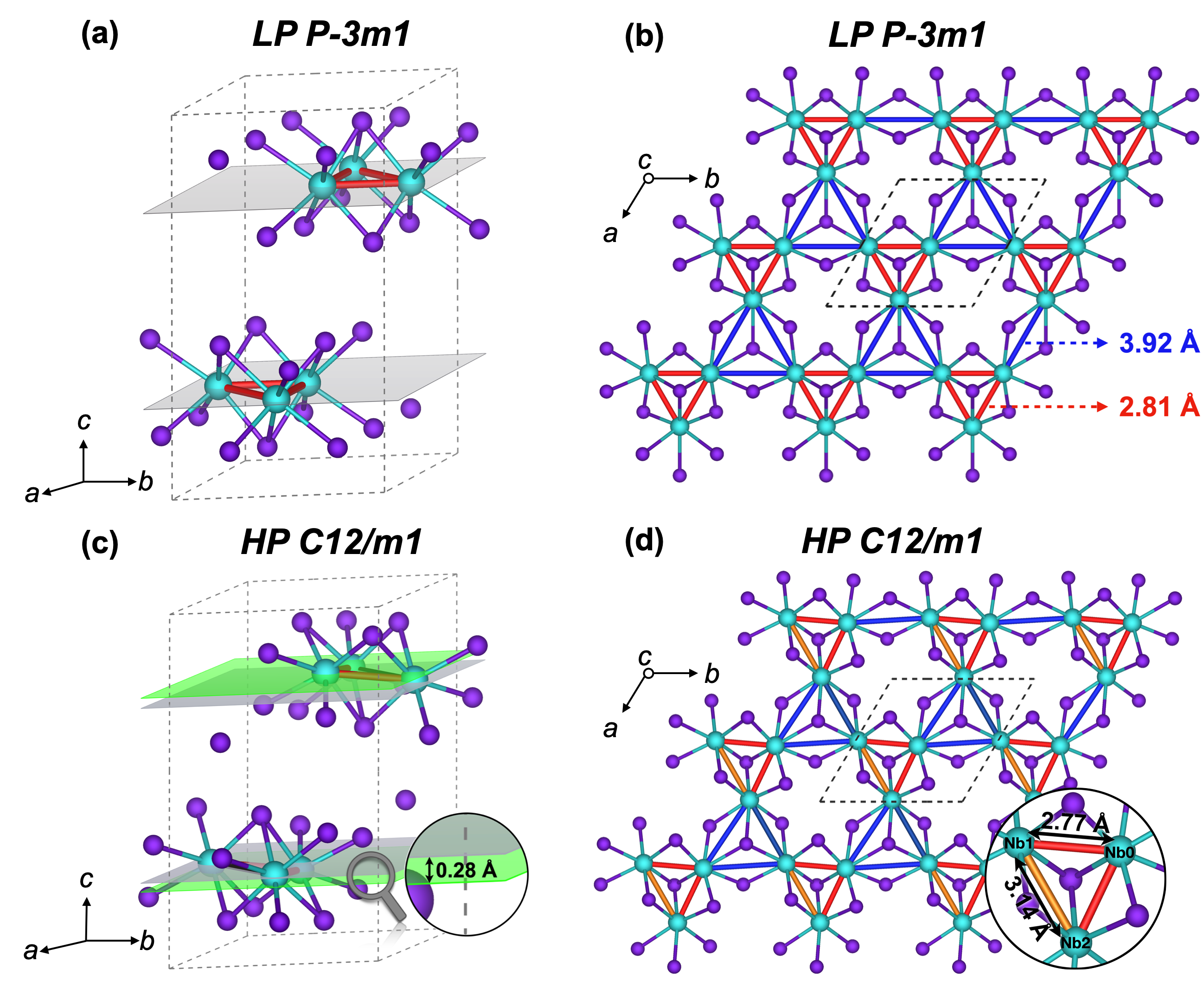}
\caption{\label{Fig1:Structure} 
\textbf{Crystal structures of Nb$_3$Cl$_8$ at low-pressure (LP) and high-pressure (HP).}  
(a) Unit cell of the LP phase ($P\overline{3}m1$). 
The gray plane perpendicular to [001] direction passes through all three Nb atoms, i.e., they are coplanar at LP. 
(b) The top view of monolayer Nb$_{3}$Cl$_{8}$ displays a breathing-kagome structure. The red triangles have equal side lengths of 2.81 Å, while the larger blue triangles have a equal side lengths of 3.92 Å.  
(c) Unit cell of the HP phase ($C12/m1$). The three Nb atoms are no longer coplanar at HP. Instead, two planes have to be introduced to include all three Nb atoms, i.e., the green and gray planes.  They displace by 0.28 Å.  
(d) Similar to (b) but for the HP structure, which shows a distorted breathing-kagome structure. The smaller triangles become isosceles, with side lengths of 2.77 Å (red) and a base length of 3.14 Å (orange).
	}
\end{figure}

\subsection{Pressure-induced local symmetry breaking}
The layered van der Waals material Nb$_3$Cl$_8$ crystalizes in $\alpha$-phase with $P\bar{3}m1$ symmetry at ambient pressure and high temperature~\cite{kennedy_experimental_1996, sheckelton_rearrangement_2017, haraguchi_magneticnonmagnetic_2017}, as illustrated in Fig.~\ref{Fig1:Structure}(a, b). 
Two layers of  Nb$_3$Cl$_{13}$ unit stack along $c$ direction with an in-plane lateral shift and are related by inversion symmetry. 
In each plane, Nb atoms form a distorted kagome lattice. 
Compared to the perfect kagome lattice which has uniform nearest neighbor Nb-Nb bond length, the distorted kagome lattice consists of big (blue) and small (red) Nb$_{3}$ triangles as shown in Fig.~\ref{Fig1:Structure}(b). 
The Nb-Nb bond lengths are 3.92~$\text{\AA}$ and 2.81~$\text{\AA}$, respectively. 
In the small Nb$_{3}$ triangles, atomic states hybridizes much more pronouncedly than in the large triangles leading to the formation of molecular states. 
In the following, we refer these small Nb$_{3}$ triangles as Nb$_{3}$ trimers. 
The three Nb atoms within each Nb$_{3}$ trimer are equivalent at ambient pressure. 
They are related by local $C_{3v}$ symmetry. 
Three Nb atoms with $[5s^14d^4]$ configuration lose 8 electrons to Cl ions resulting in 7 electrons occupying each Nb$_{3}$ trimer. 
Local crystal field splits Nb-$d$ orbital into effective $t_{2g}$ and $e_{g}$ with the latter being completely empty.
The 9 $t_{2g}$ orbitals of Nb$_{3}$ trimer hybridize yielding 9 molecular orbitals.
6 out of the 7 electrons occupied the lowest three molecular orbitals and the seventh electron makes one molecular orbital half-filled, i.e. a metallic ground state is expected.

Under pressure, the Nb$_{3}$ trimer is distorted. 
Fig.~\ref{Fig1:Structure} (c, d) displays the primitive cell and monolayer lattice structure of Nb$_3$Cl$_8$ under 9.7 GPa. 
Compared to the ambient structure in Fig.~\ref{Fig1:Structure}(a, b), there are two notable changes. 
First, the three Nb atoms in a Nb$_{3}$ trimer is no longer coplanar.
One Nb atom displaces vertically along $c$-axis from the other two by 0.28~\AA, see the green and grey planes in Fig.~\ref{Fig1:Structure}(c). 
The second change, which breaks local $C_{3v}$ symmetry, is that the three Nb-Nb bonds become inequivalent. 
Specifically, two bonds (Nb0 - Nb1 and Nb0 - Nb2) shrink from 2.81~\AA~  to 2.77~\AA~  and the third bond (Nb1 - Nb2) elongates to 3.14~\AA.
The local point symmetry is reduced to  $C_s$ and the space group becomes monoclinic $C12/m1$~\cite{jiang_pressure-driven_2022, shan_pressure-induced_2023}. 
The local symmetry breaking is expected to modify the character of molecular orbitals, but the electron occupancy and the metallic ground are expected to be unchanged.

\subsection{Electronic Structure}
\begin{figure}
\includegraphics[width=1.0\linewidth]{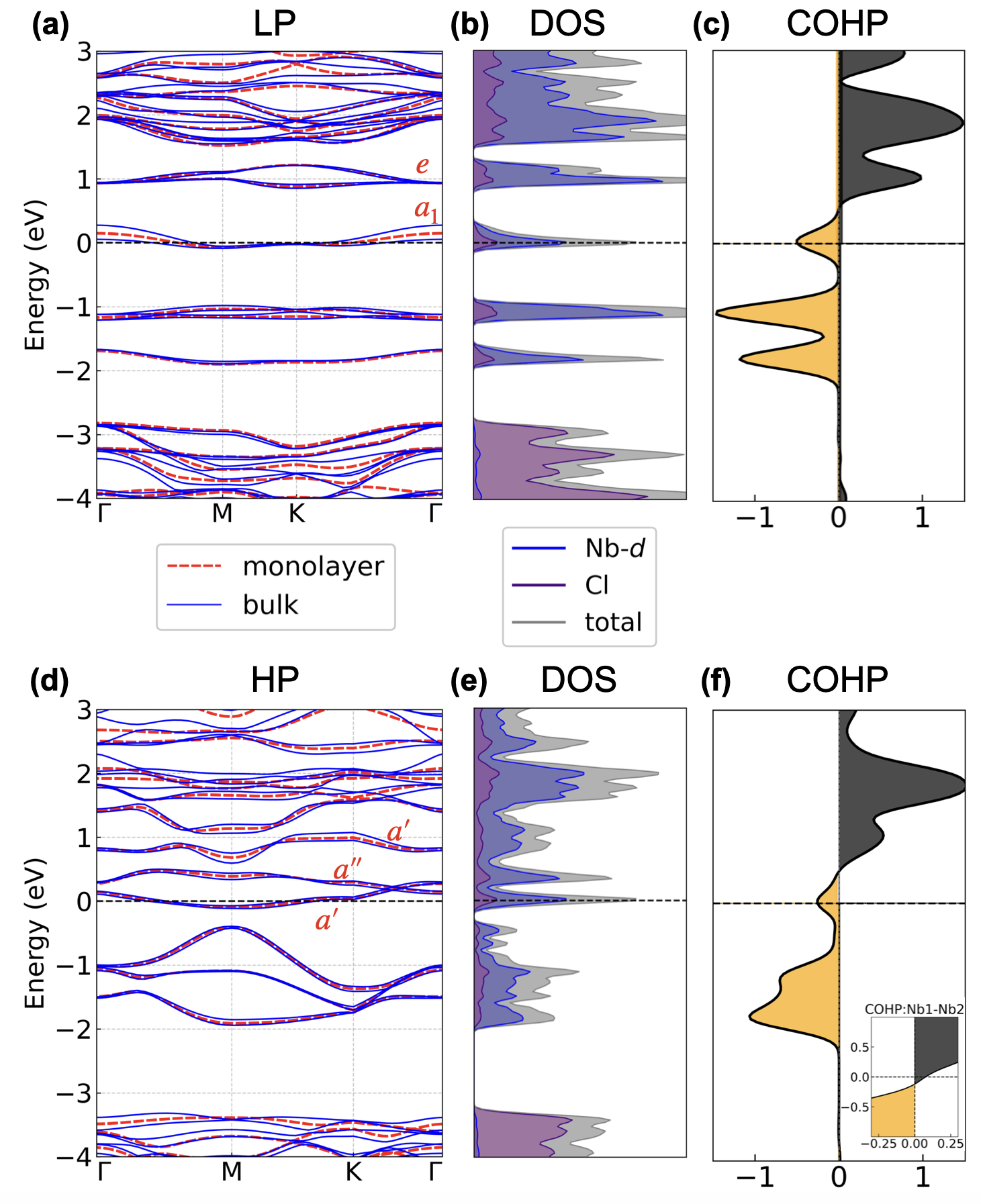}
\caption{ 
\textbf{Electronic structures of Nb$_3$Cl$_8$ at LP and HP phases. } 
(a) Band structures of the monolayer (red dashed line) and bulk (solid blue line) Nb$_3$Cl$_8$ at LP.  
(b) Projected density of states (PDOS) of bulk Nb$_{3}$Cl$_8$. Blue and purple correspond to the PDOS of Nb $d$-orbitals and Cl, respectively. The grey region represents the total density of states.  
(c) Crystal Orbital Hamiltonian Population (COHP) between Nb0-Nb1/Nb2 at LP. Bonding states (COHP $<$ 0) and antibonding states (COHP $>$ 0) are shown in gold and black, repsectively. 
(d - f) Same as (a - c) but for the HP structure.  
(f) The inset additionally shows the COHP of Nb1-Nb2 at HP.
	}
\label{Fig2:DFT}
\end{figure}

The local symmetry breaking results in a notable change in electronic structure. 
Figure~\ref{Fig2:DFT} (a, d) shows the comparison of Nb$_{3}$Cl$_{8}$ electronic structure at ambient pressure and $P = 9.7$ GPa, respectively.
They are calculated in density-functional theory (DFT) as implemented in Vienna Atomic Simulation Pacakge (VASP)~\cite{VASP-1996a, VASP-1996b}. 

Both the monolayer (red dashed line) and bulk (blue solid line) electronic structure are shown. 
Electronic structure of Nb$_{3}$Cl$_{8}$ is mainly characterized by the flat bands in energy range [-2.5, 1.5] eV~\cite{khvorykh_niobium_1995,kennedy_chemical_1991, kennedy_experimental_1996,sheckelton_rearrangement_2017, haraguchi_magneticnonmagnetic_2017, hu_correlated_2023, grytsiuk_nb3cl8_2024, gao_discovery_2023,miller_solid_1995, pasco_tunable_2019, aretz_strong_2025}.  
While the correspondence of Bloch bands in the two cases can still be identified, the band curvatures have been significantly changed. 
In particular, the flat bands in [-2, 1.5] eV in Fig.~\ref{Fig2:DFT} (a) become more dispersive in Fig.~\ref{Fig2:DFT} (d) due to the breaking of local symmetry. 
Lifting of band degeneracy is also evident in this plot. 

While the band curvature and degeneracy are changed significantly, the orbital and chemical character of these bands remain unaffected by pressure. 
Figure~\ref{Fig2:DFT} (b) and (e) show the atomic-resolved density of states (DOS). 
At LP, states below and above -2 eV are dominated by Cl and Nb, respectively. 
This characteristic is not affected by pressure. However, the sharp DOS derived from the flat bands at LP become much more broad at HP, consistent with the dispersive band structure.
We also find that the local symmetry breaking does not change the chemical nature of Nb$_{3}$Cl$_{8}$. 
As evident from the crystal orbital Hamiltonian population (COHP) analysis \cite{deringer_crystal_2011, nelson_lobster_2020} for the two short bonds (Nb0-Nb1 and Nb0-Nb2) shown in Fig.~\ref{Fig2:DFT} (c) and (f), the flat bands between [-2, -0.5] eV are derived from bonding state, while the other two flat bands between [0.5, 1.5] eV are derived from anti-bonding states.
Such a distinct chemical nature of the six flat bands remain unchanged under pressure. 
As seen from the comparison of Fig.~\ref{Fig2:DFT} (c) and (f), the chemical bonding of Nb$_{3}$Cl$_{8}$ under pressure is qualitatively same as in ambient pressure. 
\subsection{Low-energy effective model}
The pressure-induced local symmetry breaking and electronic structure change of Nb$_{3}$ trimer can be qualitatively understood from a simple tight-binding model. 
 Neglecting the six electrons in the three fully occupied molecular orbitals, we only consider the unpaired electron in a Nb$_{3}$ trimer and construct its eigenstates.  
 Each Nb atom contributes one atomic orbital (AO) denoted as $\ket{w_1}$, $\ket{w_2}$, and $\ket{w_3}$, based on which the Hamiltonian matrix of the unpaired electron can be written as
\begin{equation}\label{Eq:c3v}
\hat{H}_{\text{LP}} = -\begin{pmatrix}
0 & t & t \\
t & 0 & t \\
t & t & 0
\end{pmatrix}
+ \mu\cdot \hat{\mathbb{I}}
,
\end{equation}
where $\hat{\mathbb{I}}$ is the identity matrix, $t$ represents the hopping amplitude between three equivalent Nb sites within a trimer, and $\mu$ is the on-site energy. 
Equation~(\ref{Eq:c3v}) respects the $C_{3v}$ symmetry of Nb$_{3}$ trimer. 
Diagonalizing it yields the following eigenvalues and eigenstates
\begin{equation}
\begin{aligned}
E_{1} &= \mu-2t,\ \ \ \ \ \ \ \ &&\ket{a_1}=(\ket{w_1}+\ket{w_2}+\ket{w_3})/\sqrt{3},\\
E_{2} &=\mu+t,   &&\ket{e_1}=(-\ket{w_1}+\ket{w_3})/\sqrt{2},\\
E_{3} &=\mu+t, &&\ket{e_2} = (-\ket{w_1}+\ket{w_2})/\sqrt{2}.\\
\end{aligned}
\end{equation}
Here, $\ket{e_1}$ and $\ket{e_2}$ are two degenerate states corresponding to the two $e$ bands in Fig.~\ref{Fig2:DFT} (a), while $\ket{a_1}$ is fully symmetric, corresponding to the bonding  $a_1$ band. 
Here, one should focus on the degeneracy of bands at $\Gamma$ point, as Eq.~(\ref{Eq:c3v}) completely neglects the momentum dependence.  

Upon applying pressure, the local symmetry reduces from $C_{3v}$ to $C_{s}$, where only a mirror symmetry survives.  
To respect this symmetry change, we need to modify the hopping amplitude correspondingly. 
Suppose the basis function on Nb0 atom is $|w_1\rangle$, and they are $|w_2\rangle$ and $|w_{3}\rangle$ on Nb1 and Nb2, the Hamiltonian matrix becomes
\begin{equation}
\hat{H}_{\text{HP}} = 
-
\begin{pmatrix}
0 & t & t \\
t & 0 & t^\prime \\
t & t^\prime & 0
\end{pmatrix}
+ 
\begin{pmatrix}
\mu & 0 & 0 \\
0 & \mu^\prime & 0 \\
0 & 0 & \mu^\prime
\end{pmatrix}
,
\end{equation}
$t$ and $t^\prime$ correspond to the hopping amplitudes for Nb0-Nb1/Nb2 and Nb1-Nb2 bonds, respectively.  
$\mu$ and $\mu^\prime$ are the on-site energies for Nb0 and Nb1 (Nb2).
The corresponding eigenvalues are
\begin{align}
E_{1} & = \frac{1}{2}[-\sqrt{(\mu - \mu^\prime - t^\prime)^2 + 8t^2} + \mu + \mu^\prime + t^\prime]\;, \nonumber \\
E_{2} & = \mu^\prime + t^\prime\;, \nonumber \\
E_{3} & = \frac{1}{2}[\sqrt{(\mu - \mu^\prime - t^\prime)^2 + 8t^2} + \mu + \mu^\prime + t^\prime]\;.
\end{align}
All three eigenvalues are different when $\mu\ne\mu^\prime$ and $t\ne t^\prime$, i.e. the degeneracy of the two $e$ bands is lifted by the local symmetry breaking. 

\begin{figure}[htbp]
\includegraphics[width=1.0\linewidth]{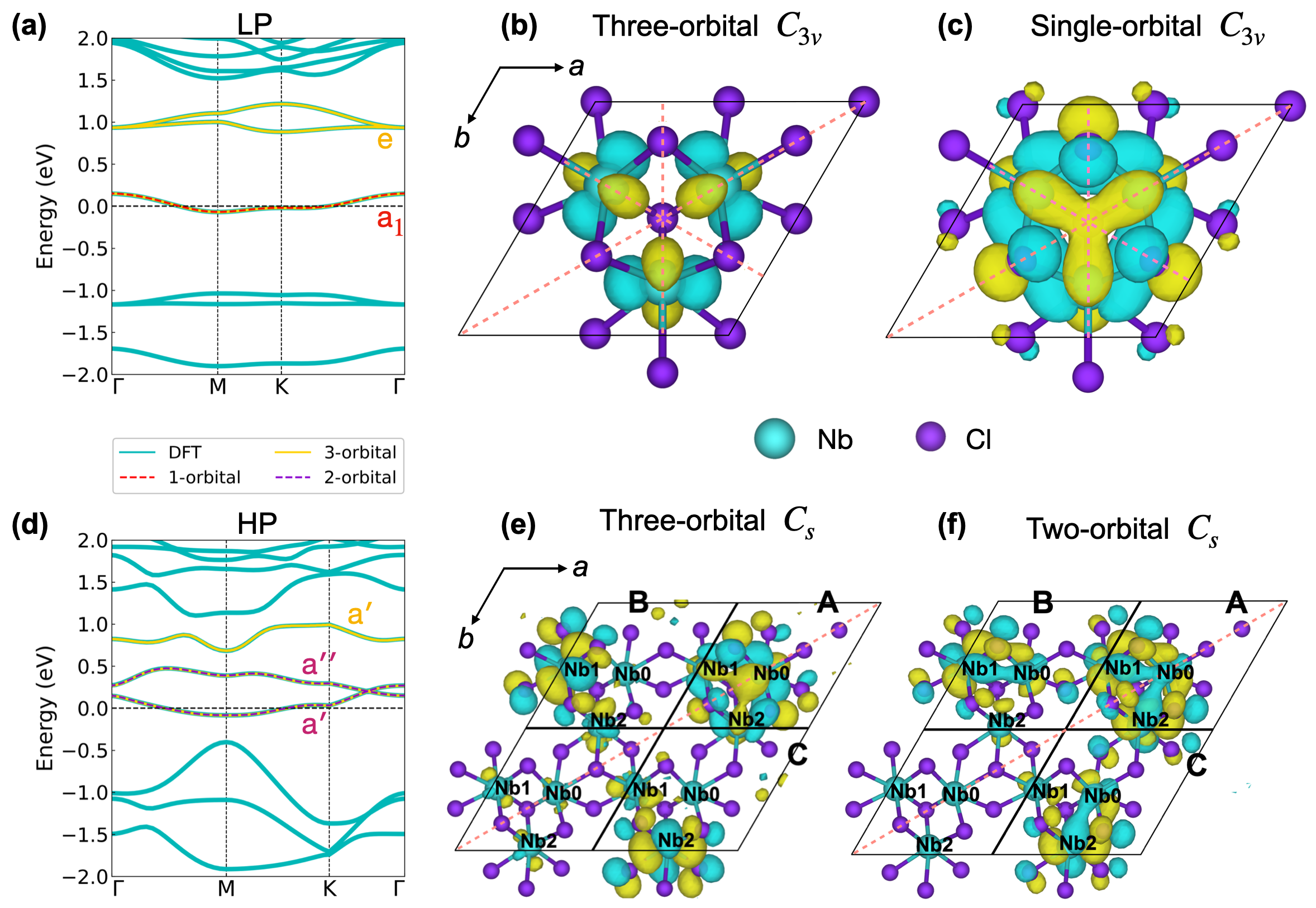}
\caption{
\textbf{Tight-binding model construction for the monolayer Nb$_{3}$Cl$_{8}$.}
(a) Comparison of the tight-binding model dispersions to the DFT band structure. The red dashed line and the yellow solid lines corresponds to a three-orbital model and a single-orbital model, respectively.
 (b, c) The real-space wannier functions of the single- and three-orbital models, where the red dashed line denotes the three mirror planes.
 (d) Same as (a) but for the HP structure. The red dashed lines and the yellow solid lines correspond to the two-orbital and three-orbital models, respectively.
 (e, f) The real-space wannier functions for the three-orbital and two-orbital models, in which there is only one mirror plane (red dashed line). 
 While the wannier centers are still on Nb sites,  wannier functions at HP phase are much extended in real-space. To better visualize them, we plot each wannier function in an individual cell A, B, and C. 
In (f), in addition to the two wannier functions plotted in cell B and C, we also show their sum in cell A which clearly displays only one mirror symmetry.
}
\label{Fig3:Wannier}
\end{figure}

After qualitatively understanding the band structure change induced by the local symmetry breaking of Nb$_{3}$ trimer, we now construct more accurate tight-binding models for a few low-energy DFT bands, which will be used as the starting point for the study of the interaction-induced charge gap discussed in the next section. 
For simplicity, here we only take the monolayer Nb$_{3}$Cl$_{8}$ at LP and HP as examples.
The downfolding of wannier tight-binding model for the bulk is technically same as for the monolayer, but the required number of orbitals are doubled. 

We employ the maximal localization wannier function technique~\cite{W90-2020} to construct a three-orbital model for both the LP and HP monolayer Nb$_{3}$Cl$_{8}$. 
Figure~\ref{Fig3:Wannier} displays the comparison of DFT and wannier band structures. They nicely agree with each other. 
The real-space wannier functions are shown in Fig.~\ref{Fig3:Wannier} (b) and (e). 
The centers of the three AOs locate at the three Nb sites and they nicely respect the $C_{3v}$ and $C_s$ local symmetry for the LP and HP structure. 
Notably, the  the three wannier functions in Fig.~\ref{Fig3:Wannier} (e) are mirror-symmetric with respect to the (11) direction but don't have any other symmetry. 
Three-orbital model is the minimal model that maintains the AO center at each Nb site. 
While further downfolding to a smaller tight-binding model is possible, the atomic character of the wannier functions will be lost. 
Figure~\ref{Fig3:Wannier} (c) shows the wannier function of single-orbital tight-binding model for band $a_1$ in Fig.~\ref{Fig3:Wannier} (a), whose center is in the middle of Nb$_{3}$ trimer and respects the full $C_{3v}$ symmetry, demonstrating a clear molecular shape. 
Similarly, we can also downfold $a^\prime$ and $a^{\prime\prime}$ in Fig.~\ref{Fig3:Wannier} (d) to a two-orbital model. 
The corresponding wannier functions are shown in Fig.~\ref{Fig3:Wannier} (f).
 Note that, under pressure, the real-space wannier functions become more extended. To better visualize them, in Fig.~\ref{Fig3:Wannier}(e) and (f) we plot each wannier function in an individual cell (labeled as A, B, and C). For the two-orbital model, in addition to the two wannier functions shown in cell B and C, we also plot them together in cell A. 
 Both the wannier functions in Fig.~\ref{Fig3:Wannier} (c) and (f) demonstrate a clear molecular character. 
We, thus, can expect a cluster Mott state for the HP Nb$_{3}$Cl$_{8}$ as well. 

\subsection{Cluster Mott state at both LP and HP phases}

% FIGURE
\begin{figure}[t]
\includegraphics[width=1.0\linewidth]{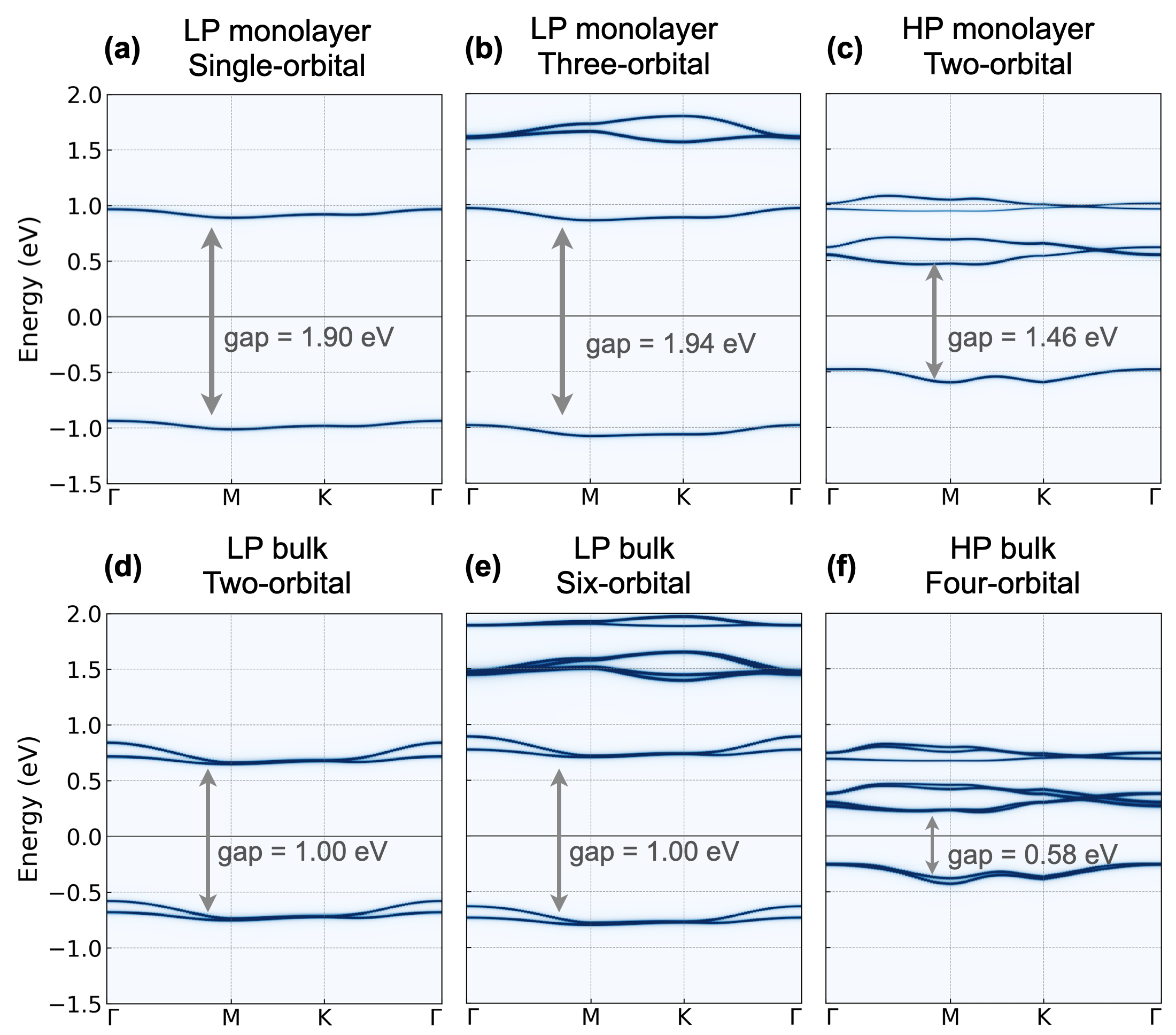}
\caption{
\textbf{Interacting spectral function as calculated from DFT + DMFT:}
 (a) The charge gap is estimated as 1.9 eV for the monolayer Nb$_{3}$Cl$_8$ at LP which becomes 1.0 eV for the (d) bulk.
 (b) Three-orbital model gives rise to a gap value of 1.94 eV, which is consistent with that obtained in (a).   
 Similarly, in (e) the charge gap of the bulk estimated from the six-orbital model is same as that in (d).  
 (c, f)  At HP, the enhanced bandwidth and reduced Coulomb interactions shrink the charge gap to (c) 1.46 eV and (f) 0.58 eV for the monolayer and bulk system, respectively. 
 In these two calculations, a two-orbital and a four-orbital TB model for the monolayer and the bulk are used. 
	}
\label{Fig4}
\end{figure}

\begin{table*}[htbp]
\centering
% \captionsetup{width=\linewidth}
\caption{The strength of the density-density Coulomb repulsions at LP and HP phases of Nb$_3$Cl$_8$ as evaluated by using cRPA based on the different TB models discussed in Fig.~\ref{Fig3:Wannier}. 
$U_0$, $U_1$, and $U_2$ correspond to the local Coulomb interactions at Nb0, Nb1, and Nb2 sites. 
$V^{\text{in}}_{mn}$ denotes the non-local interactions between two Nb sites. 
For the bulk system, the interlayer Coulomb interaction strength is additionlly given as $V^{\text{out}}$.}
\renewcommand{\arraystretch}{1.1} % 调整行高
\setlength{\tabcolsep}{14pt} % 调整列间距
\begin{tabular}{ccccccccc}
\toprule
\toprule

 & & \multicolumn{2}{c}{LP\ mono} & \multicolumn{2}{c}{LP\ bulk} & \multicolumn{2}{c}{HP\ mono} & HP\ bulk \\
\cmidrule(r){3-4} \cmidrule(r){5-6} \cmidrule(r){7-8} \cmidrule(r){9-9}
 & (eV) & 1-orb & 3-orb & 2-orb & 6-orb & 2-orb & 3-orb & 4-orb \\
\midrule
\multirow{3}{*}{on-site} 
    & $U_{0}$ & 1.90 & 2.63 & 1.40 & 2.17 & ---  & 1.17 & ---  \\
    & $U_{1}$ & ---  & 2.63 & ---  & 2.17 & 1.45 & 1.62 & 0.97 \\
    & $U_2$   & ---  & 2.63 & ---  & 2.17 & 1.45 & 1.62 & 0.97 \\

\midrule
\multirow{4}{*}{inter-site}
    & $V_{01}^{\text{in}}$ & --- & 1.72 & ---   & 1.29 & ---  & 1.04 & ---\\
    & $V_{02}^{\text{in}}$ & --- & 1.72 & ---   & 1.29 & ---  & 1.04 & ---\\
    & $V_{12}^{\text{in}}$ & --- & 1.72 & ---   & 1.29 & 1.06 & 1.07 & 0.61\\
    & $\  V^{\text{out}}$  & --- & ---  & 0.25  & 0.28 & ---  & ---  & 0.18\\
    
\bottomrule
\bottomrule
\end{tabular}

\label{tabII:cRPA}
\end{table*}

The flat bands at LP significantly enhance the effect of Coulomb interaction, in particular the non-local interaction within the Nb$_{3}$ trimer~\cite{hu_correlated_2023}. 
The experimentally observed insulating states~\cite{nanolett_feng_2022, gao_discovery_2023, jiang_pressure-driven_2022} together with the theoretical understanding~\cite{hu_correlated_2023} convincingly reveal a cluster Mott ground state. 
The pressure-induced local symmetry breaking enhances the bandwidth and lifts the band degeneracy of these flat bands at HP. 
While our intuitive understanding discussed before favors the persistence of the cluster Mott state at HP, a more rigorous and direct evidence is still lacking. 
In the following, by employing DFT + dynamical mean-field theory (DMFT)~\cite{DMFT-Georges1996} we will show that the insulating ground state at HP~\cite{jiang_pressure-driven_2022} is indeed a cluster Mott state but with a reduced charge gap. 

To this end, we consider the following interacting Hamiltonian
\begin{equation}
H = -\sum_{ij, \sigma} t_{ij} c_{i\sigma}^\dagger c_{j\sigma} + \sum_{i} U_i n_{i\uparrow} n_{i\downarrow} 
+ \frac{1}{2} \sum_{i \neq j, \sigma, \sigma'} V_{ij} n_{i\sigma} n_{j\sigma'},
\end{equation}
where $c_{i\sigma}^\dagger$ creates an electronic state at site $i$ with spin $\sigma$. 
$U_i$, $V_{ij}$ correspond to the on-site and inter-site Coulomb repulsions. 
We theoretically estimate these interaction parameters for different TB models by using the $\textit{ab initio}$ constrained Random Phase Approximation (cRPA) method~\cite{CRPA-Aryasetiawan1998, CRPA-Aryasetiawan2004, crpa}.  
The results are shown in Tab.~\ref{tabII:cRPA}. 
Consistent with the characteristic of a cluster Mott insulator, in both LP and HP phases, the strength of the intra-trimer, inter-site repulsion ($V^\text{in}$) is comparable to the on-site repulsion ($U$). 
In contrast, the inter-layer Coulomb repulsion ($V^{\text{out}}$) is negligible, and the inter-trimer repulsion is also expected to be small.
Furthermore, we also observe that the corresponding interaction parameters are larger in monolayer than in bulk systems, consistent with the fact that there is less electronic screening in monolayer Nb$_{3}$Cl$_{8}$ than in the bulk.

With the TB models and the corresponding Coulomb parameters, we are now ready to present the interacting spectral function and characterize the nature of the ground state at HP. 
We solved the DMFT equation with our home-made \textit{Package for Analyzing Correlated Systems} (PACS) and employed the Hubbard-I as impurity solver~\cite{PhysRevB.85.115103, PACS_1}. 
Figure~\ref{Fig4} displays the DMFT spectral functions for both monolayer and bulk Nb$_{3}$Cl$_{8}$ under LP and HP. 
Corresponding to the TB models downfolded in Fig.~\ref{Fig3:Wannier}, Fig.~\ref{Fig4} (a) shows the interacting spectral function of the single-orbital TB model. 
One unpaired electron occupies this band leading to a standard half-filled Hubbard model which develops lower and upper Hubbard bands under local interaction $U$. 
With the estimated Coulomb parameters in Tab.~\ref{tabII:cRPA}, the charge gap is obtained as 1.9 eV. 
Similarly, in the three-orbital model where the atomic orbital character of each basis function is maintained, the same charge gap is obtained although two completely different sets of Coulomb parameters are used for the two models. 
The nice agreement between the molecular description with single-orbital model and the atomic description with the three-orbital model and non-local interactions strongly confirm the persistent cluster Mott states at HP in Nb$_{3}$Cl$_8$. 
Similarly, a HP, the charge gap is estimated as 1.46 eV which is smaller than that in LP phase. 
The reduced but finite charge gap is consistent with the experimental observation of the insulating ground state of Nb$_{3}$Cl$_{8}$ under pressure~\cite{jiang_pressure-driven_2022}. 
The molecular character in Fig.~\ref{Fig3:Wannier} (f), the amplitude of the non-local interaction in Tab.~\ref{tabII:cRPA}, and the finite charge gap in Fig.~\ref{Fig4} convincingly demonstrate that the ground state of the HP Nb$_{3}$Cl$_{8}$ is still a cluster Mott insulator. 
While the pressure-induced symmetry breaking reduces the global symmetry and lifts band degeneracy, it does not change the molecular nature of the band at the Fermi level. 
The increased bandwidth and the electronic screening reduces the effect of Coulomb interaction, resulting in a reduced charge gap. 
Our work, thus, provides a consistent explanation to the recently observed insulating ground state of Nb$_{3}$Cl$_{8}$ with a smaller gap under pressure.

\section{Conclusion}

In this work, we theoretically study the electronic structure of Nb$_{3}$Cl$_{8}$ under pressure and explain the origin of the reduced charge gap recently observed in experiment~\cite{jiang_pressure-driven_2022}. 
Our study reveals a pressure-induced local symmetry breaking which reduces the symmetry of Nb$_{3}$ trimer from $C_{3v}$ to $C_{s}$. 
The local symmetry breaking significantly enhances the bandwidth and lifts the band degeneracy. 
However, we find that the ground state of HP phase is still a cluster Mott insulator with considerably large non-local Coulomb interactions. 
This conclusion is supported by the modular character of the wannier functions, the \textit{ab initio} evaluation of the Coulomb parameters, and the calculation of the correlated charge gap. 
The enhanced bandwidth and the reduced Coulomb interactions at HP shrink but does not close the charge gap, consistent with the experimental observation. 
The presistance of cluster Mott insulating state under structural distortion highlights the remarkable resilience of the correlated insulating state in Nb$_3$Cl$_8$, suggesting external pressure to be an effective knob to continuously tune the cluster Mott gap, a property that has strong implication to optical device application. 
These findings establish a theoretical foundation for understanding and potentially controlling emergent quantum states in transition metal cluster compounds through symmetry engineering.

\begin{acknowledgments}
This work was supported by the National Key R\&D Program of China (No. 2022YFA1402703), Sino-German Mobility program (No. M-0006), and Shanghai 2021- Fundamental Research Area (No. 21JC1404700). Part of the calculations was performed at the HPC Platform of ShanghaiTech University Library and Information Services, and the School of Physical Science and Technology.
\end{acknowledgments}

\bibliographystyle{apsrev4-2}
\bibliography{ref}

\end{document}